\RequirePackage{fix-cm}
\documentclass[twocolumn,epjc3]{svjour3}  

\smartqed  
\usepackage{graphicx}
\begin{document}

\title{Enhanced light signal for the suppression of pile-up events in Mo-based bolometers for the 0$\nu\beta\beta$ decay search.}

\author{A. Ahmine\thanksref{addr7}
\and
A. Armatol\thanksref{addr3}
        \and
        I. Bandac\thanksref{addr2}
        \and
        L. Berg\'e\thanksref{addr1}
        \and
        J.M. Calvo-Mozota\thanksref{addr2,addr6}
        \and
        P. Carniti\thanksref{addr4}
        \and
        M. Chapellier\thanksref{addr1}
        \and
        T. Dixon\thanksref{addr1}
        \and  
        L. Dumoulin\thanksref{addr1}
        \and
        A. Giuliani\thanksref{addr1}
        \and
        Ph. Gras\thanksref{addr3}
        \and
        F. Ferri\thanksref{addr3}
        \and
       L. Imbert\thanksref{addr1}
        \and
       H. Khalife\thanksref{addr3}
        \and
        P. Loaiza\thanksref{addr1}
                \and
       P. de Marcillac\thanksref{addr1}
        \and
       S. Marnieros\thanksref{addr1}
        \and
       C.A. Marrache-Kikuchi\thanksref{addr1}
        \and
        C. Nones\thanksref{addr3}
        \and
        E. Olivieri\thanksref{e1,addr1}
        \and
        A. Ortiz de Sol\'orzano\thanksref{addr8}
        \and 
        G. Pessina\thanksref{addr4}
        \and
        D.V. Poda\thanksref{addr1}
        \and
        Th. Redon\thanksref{addr1}
        \and
       J.A. Scarpaci\thanksref{e2,addr1}
       \and
       M. Vel\'azquez\thanksref{addr7}
       \and
       A. Zolotarova\thanksref{addr3}
}

\thankstext{e1}{e-mail: emiliano.olivieri@ijclab.in2p3.fr}
\thankstext{e2}{e-mail: jean-antoine.scarpaci@ijclab.in2p3.fr}


\institute{Universit\'e Grenoble Alpes, CNRS, Grenoble INP, SIMAP 38402 Saint Martin d'H\'eres, France\label{addr7}
\and
IRFU, CEA, Universit\'e Paris-Saclay, F-91191 Gif-sur-Yvette, France \label{addr3}
                      \and
                      Laboratorio Subterr\'aneo de Canfranc, 22880 Canfranc-Estaci\'on, Spain \label{addr2}
           \and
           Universit\'e Paris-Saclay, CNRS/IN2P3, IJCLab, 91405 Orsay, France\label{addr1}
           \and
          Escuela Superior de Ingeniería y Tecnología, Universidad Internacional de La Rioja, 26006 Logroño, Spain \label{addr6}
           \and
           INFN, Sezione di Milano Bicocca, I-20126 Milano, Italy\label{addr4}
           \and
           Centro de Astropart\'iculas y F\'isica de Altas Energ\'ias, Universidad de Zaragoza, Zaragoza 50009, Spain\label{addr8}
           }

\date{Received: date / Accepted: date}

\maketitle

\begin{abstract}
Random coincidences of events could be one of the main sources of background in the search for neutrino-less double-beta decay of $^{100}$Mo with macro-bolometers, due to their modest time resolution.
Scintillating bolometers as those based on Li$_2$MoO$_4$ crystals and employed in the CROSS and CUPID experiments can eventually exploit the coincident fast signal detected in a light detector to reduce this background. However, the scintillation provides a modest signal-to-noise ratio, making difficult a pile-up pulse-shape recognition and rejection at timescales shorter than a few ms. Neganov-Trofimov-Luke assisted light detectors (NTL-LDs) offer the possibility to effectively increase the signal-to-noise ratio, preserving a fast time-response, and enhance the capability of pile-up rejection via pulse shape analysis.
In this article we present: a) an experimental work performed with a Li$_2$MoO$_4$ scintillating bolometer, studied in the framework of the CROSS experiment, and utilizing a NTL-LD; b) a simulation method to reproduce, synthetically, randomly coincident two-neutrino double-beta decay events; c) a new analysis method based on a pulse-shape discrimination algorithm capable of providing high pile-up rejection efficiencies. We finally show how the NTL-LDs offer a balanced solution between performance and complexity to reach background index \break $\sim$$10^{-4}$ counts/keV/kg/year with 280~g Li$_2$MoO$_4$ ($^{100}$Mo enriched) bolometers at 3034 keV, the Q$_{\beta\beta}$ of the double-beta decay, and target the goal of a next generation experiment like CUPID.
\end{abstract}

\section{Introduction}
The observation of the neutrino-less double-beta ($0\nu2\beta$) decay would imply the violation of lepton number conservation and establish the Majorana nature of neutrino \cite{Majo:1937,Raca:1937,Pont:1968,Sche:1980}. Cryogenic bolometers are very competitive detectors to search for this extremely rare process ($T_{1/2}^{0\nu2\beta} > $ 10$^{24}$--$10^{26}$ yr \cite{Arnold:2015,Gando:2016,Alvis:2019,Anton:2019,Azzolini:2019,Adams:2020,Armengaud:2021,Augier:2022,Agostini:2020}) in a few theoretically and experimentally favorable nuclei. The lithium molybdate compound (Li$_2$MoO$_4$) has been experimentally demonstrated as one of the most promising materials to this end \cite{Bekker:2016,Armengaud:2017,Armengaud:2020a,Armengaud:2021,Augier:2022,Poda:2021}, thanks to the recent progress in the techniques to synthesize large, high-quality $^{100}$Mo-enriched radio-pure scintillating single crystals \cite{Chernyak:2017,Augier:2022,Armatol:2021a,Armengaud:2017,Grigo:2017}, and providing excellent bolometric performances overall \cite{Armengaud:2020a}. 

Both CROSS \cite{Bandac:2020} and CUPID \cite{CUPID:2019} experiments develop a technology based on Li$_2$MoO$_4$ macro-bolometers coupled to bolometric light detectors and use neutron-transmutation-doped Ge thermistors (NTD) \cite{haller:1994} as sensors. These sensors can be produced in large numbers and they can be read out via simple, conventional, low-noise JFET-based electronics \cite{Pessina:electronics}. Their major drawback is an intrinsic slow (1--10 ms) response time, essentially due to 1) their high impedance at the optimal working point, 2) the gluing interface with the crystal, and 3) the internal electron-phonon decoupling. Slow response time can lead to pile-up events mimicking single, normal events at the Q-value of the double-beta decay $Q_{\beta\beta}$ \cite{Chernyak:2014,Chernyak:2017}. The identification of these events is extremely important when searching for $0\nu2\beta$ decay of $^{100}$Mo: in fact, having the $^{100}$Mo a relatively "short" two-neutrino double-beta ($2\nu 2\beta$) half-life ($T_{1/2}^{2\nu2\beta} = 7.1\cdot10^{18}$ years \cite{Armengaud:2020b}), this latter can bring a large background of pile-up events in the ROI.

Other experiments as AMoRE \cite{Alenkov:2015}, also looking for $^{100}$Mo neutrinoless double-beta decay, adopted metallic-magnetic-calorimeter (MMC) as temperature sensors \cite{Kempf:2018}. The major advantage of this sensor technology is to provide signals with faster response times than that of the NTDs, reducing the pile-up background in the ROI.

The counting rate of randomly coincident $2\nu2\beta$ decay events in a 280\ g Li$_2${}$^{100}$MoO$_4$ CUPID detector is estimated on the level of $3.3\cdot 10^{-4}$ counts/keV/kg/year (ckky) at 3034 keV, assuming a time-resolving capability of 1 ms \cite{Chernyak:2017}: randomly-coincident $2\nu2\beta$ decay events represent at the moment the main source of background for the next-generation, large-scale, high-radio-pure experiments based on the CUPID technologies \cite{CUPID:2019,Armatol:confproc}.

Pile-up rejections using pulse-shape analysis of the heat channel have been presented \cite{Armatol:2021b}: rejection performances of 90\% for pile-up times down to 2~ms were reported. However, these values are not sufficient to reach the ultimate CUPID goal.

The approach followed here is the same as that reported in Ref. \cite{Chernyak:2017}, consisting of exploiting the light signal to improve pile-up rejection through pulse-shape discrimination. In fact, as reminded above, both CROSS and CUPID will be provided with optical bolometers to detect the scintillation light in coincidence with the heat signal measured in the Li$_2${}$^{100}$MoO$_4$ crystal. These light detectors consist of thin Ge wafers equipped with NTD's as the Li$_2${}$^{100}$MoO$_4$ crystal. The collected light is detected as a temperature pulse in the wafer. The primary function of the double heat-light readout is the rejection of the alpha background by exploiting the lower scintillation yield of alphas with respect to betas/gammas for the same deposited energy. Besides this,  the light detector can play a crucial role also to mitigate the pile-up background, since its signal rise-time is about ten time shorter than that of the heat signal. In order for this method to be effective, the signal-to-noise ratio needs to be enhanced, which can be obtained through the so-called Neganov-Trofimov-Luke (NTL) effect \cite{Neganov,Luke}.


With respect to past work, which was mainly conceptual,  our results here are supported by extensive experimental results and more sophisticated simulations, providing a convincing and operational method to reject the background related to $2\nu2\beta$ decay events down to the desired level. 

In this study we report about an experimental and simulation work which provides a technological solution combined with simulations and analysis methods to demonstrate how to reach a background index (BI) lower than $\sim$10$^{-4}$ ckky in the region of interest (ROI) of $^{100}$Mo $0\nu2\beta$ decay.

It consists of three main sections: 
\begin{itemize}
    \item[-] an experiment performed with a 245\ g Li$_2$MoO$_4$ scintillating bolometer equipped with a NTL effect boosted light detector (NTL-LD) as such in \cite{Novati:2019}. The bolometer was realized in the framework of the CLYMENE project and the experiment run at the C2U (CROSS Cryogenic Underground) facility at Canfranc Underground Laboratory (Spain) as part of the CROSS/CLYMENE joint R\&D program (Section~\ref{sec:expe});
    \item[-] a new analysis method based on an optimal-filtering pulse-shape discrimination algorithm, capable to provide enhanced pile-up rejection efficiencies (Section~\ref{sec:data});   
    \item[-] a simulation method to reconstruct synthetically produced random coincidences of $2\nu2\beta$ decays: it is fully based on signals and noise obtained from detector measurements (Section~\ref{sec:sim}).
\end{itemize}

\section{The experiment}
\label{sec:expe}
\subsection{Setup description}
\label{subsec:expsetup}

\begin{figure*}[!h]
\centering
     \includegraphics[width=0.9\textwidth]{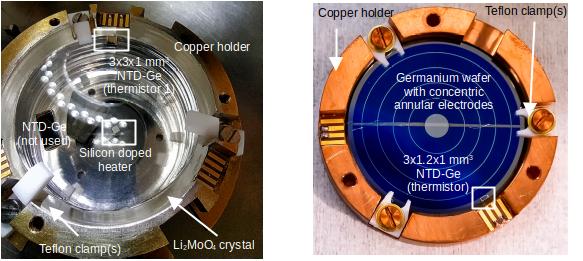}
    \caption{(LEFT) Main heat bolometer, accommodated in an silver-coated copper case and operated at 15~mK. It is equipped with 2 different NTD-Ge thermistors and a 300~k$\Omega$ silicon doped heater;  (RIGHT) a NTL-LD germanium detector, equipped with interconnected, concentric electrodes: it faces the heat bolometer and closes the cavity on top. It has an effective signal-to-noise gain of the order of 10 when the differential of potential of the two set of interconnected electrodes is kept at 50~V. An anti-reflecting SiO coating (bluish) is deposited on the germanium wafer, between the electrodes.}
    \label{fig:setup}
\end{figure*}

Figure \ref{fig:setup} shows a picture of the scintillating macro-bolometer mounting used in this work. The experiment was performed in a low radioactivity pulse tube dilution refrigerator (Hexadry-400, from Cryoconcept) at around 15~mK. The main bolometer consists of a 245~g Li$_2$MoO$_4$ cylindrical single crystal; it is fit in a silver-coated reflective cavity and equipped with two NTD-Ge sensors (3$\times$3$\times$1~mm$^3$ and  10$\times$3$\times$1~mm$^3$) and one 300~k$\Omega$ Si:P heater \cite{andreotti:2012}. This detector was developed within the CLYMENE project \cite{Velazquez:2017,stelian:2020} (ANR funded) and studied in synergy with the CROSS ERC granted project \cite{Bandac:2020}.

A Neganov-Trofimov-Luke-effect boosted germanium light detector, identical to the one described in \cite{Novati:2019}, is coupled to the cavity and records the scintillation emitted in coincidence with the particle events occurring in the main bolometer, allowing to reject the $\alpha$ background. This detector is fabricated with a 170 $\mu$m thick, $\oslash=$ 44 mm high purity Ge wafer on top of which annular aluminum  electrodes of 200 $\mu$m width and 3.8 mm pitch are evaporated. The electrodes are interconnected to produce two different electrode sets, and kept at a voltage difference $V_e$. The NTL-LD was operated with several $V_e$ up to a maximum of 50~V: for this value it returned an enhanced sensitivity about 10 times larger than at $V_e=$ 0~V and no additional contribution in the Noise Power Spectrum (NPS).


An optical fiber is coupled with a room temperatures 880\ nm wavelength LED and used to shine the NTL-LD in the cavity. The LED was driven via Keysight 33500B wave-function generator sending typically square voltage pulses of about 400\ ns time-width and a voltage amplitude around 1 V.

\subsection{Measurements}
\label{subsec:meas}

We artificially produced pile-up events in the light detector via LED photo-pulses sent through the optical fiber.
The voltage amplitudes, A1 and A2, of the wave-function generator were tuned, together with the pulse duration (typically a few $\mu$s, much shorter than the rise-time $\tau_r$ of the light detector of the order of 1~ms), to provide single pulses of energy $E$ = 450\ eV. Hence, a perfect synchronous two LED pulses ($\Delta$t${}={}$0~ms) delivers 900~eV, $i.e.$ the amount of scintillating energy recovered on a single light detector for an event at the $^{100}$Mo Q$_{\beta\beta}$ (3\ MeV), for a light yield of 0.3~ keV/MeV as in CROSS or CUPID setup \cite{Armatol:2021a,Alfonso:2022}. For simplicity, we worked with A1${}={}$A2 as the simulation presented hereafter shows that this pulse amplitude combination gives the main pile-up contributions, and sending two subsequent pulses spaced by given $\Delta$t time interval. We adopted the following $\Delta$t time pattern: 0.1; 0.3; 0.5; 0.7; 1.0; 3.0; 5.0; 0.05 ~ms. We delivered 100 (pile-up) signals for each $\Delta$t, at a repetition rate of 0.2\ Hz. We have performed several set of measurements, by varying the NTL-LD electrode voltage bias V$_e$ (10, 30 and 50~V). As a precaution, between each set of voltage we used the LED to charge-reset the NTL-LD, and clean the residual, spurious electric field  built-up in the semiconductor due to charge trapping as described in \cite{Olivieri:2009}\footnote{However a similar NTL light detector was operated in the same set-up during one month for background acquisition without need of any charge-reset procedure.}.

We measured an average performance on the light channel of $\sigma\sim90$~eV ($\sim9$~eV) baseline noise when operated at V$_e${ }={ }0~V (50~V) and sensitivity of 1~$\mu$V/keV (10~$\mu$V/keV). We explored different NTL-LD detector working points by scanning with respect to different sensor bias currents (Table \ref{tab:workingpoint}). We finally chose to work at a bias of 1~nA, which provides a signal rise-time of 1.2~ms.
\begin{table*}[]
    \centering
      \caption{Characterization of different NTL-LD working points with respect to several $I_{NTD}$ thermal sensor bias current, for a mixing chamber temperature regulated at 15\ mK. We finally performed all the measurements at 1~nA bias current, which is chosen to obtain a good compromise between fast response and decent signal amplitude.}
    \label{tab:workingpoint}
    \begin{tabular}{c|c|c|c}
         $I_{NTD}$ [nA] & $R_{NTD}$ [M$\Omega$] &Responsivity [A.U.] & $\tau_r$ (10--90 \%) [ms] \\
         \hline
         0.28 & 10 & 1&2\\
        1 & 2.3 & 0.363&1.2\\
         2.9 & 0.7 & 0.182& 0.8
    \end{tabular}

\end{table*}

The NTL-LD signals were filtered with an analog low-pass Bessel filter: the cut-off frequency was set at 500\ Hz \cite{Arnaboldi:2002,Carniti:2020,Carniti:2022}, above the intrinsic detector cut-off frequency $F_{intr}=0.35/\tau_r$. All the measurements where 24 bit sampled at a frequency of 5~kHz and continuously recorded (streaming) during the course of the experiment. The streaming files were analyzed offline via an analysis software using Gatti-Manfredi matched optimum filter \cite{Gatti:1986}: it returns the energy estimate and pile-up rejection parameters, as defined hereafter in Section~\ref{sec:data}.

\section{Data analysis: method and results }\label{sec:data}

\begin{figure}[!h]
\centering
\includegraphics[width=0.45\textwidth]{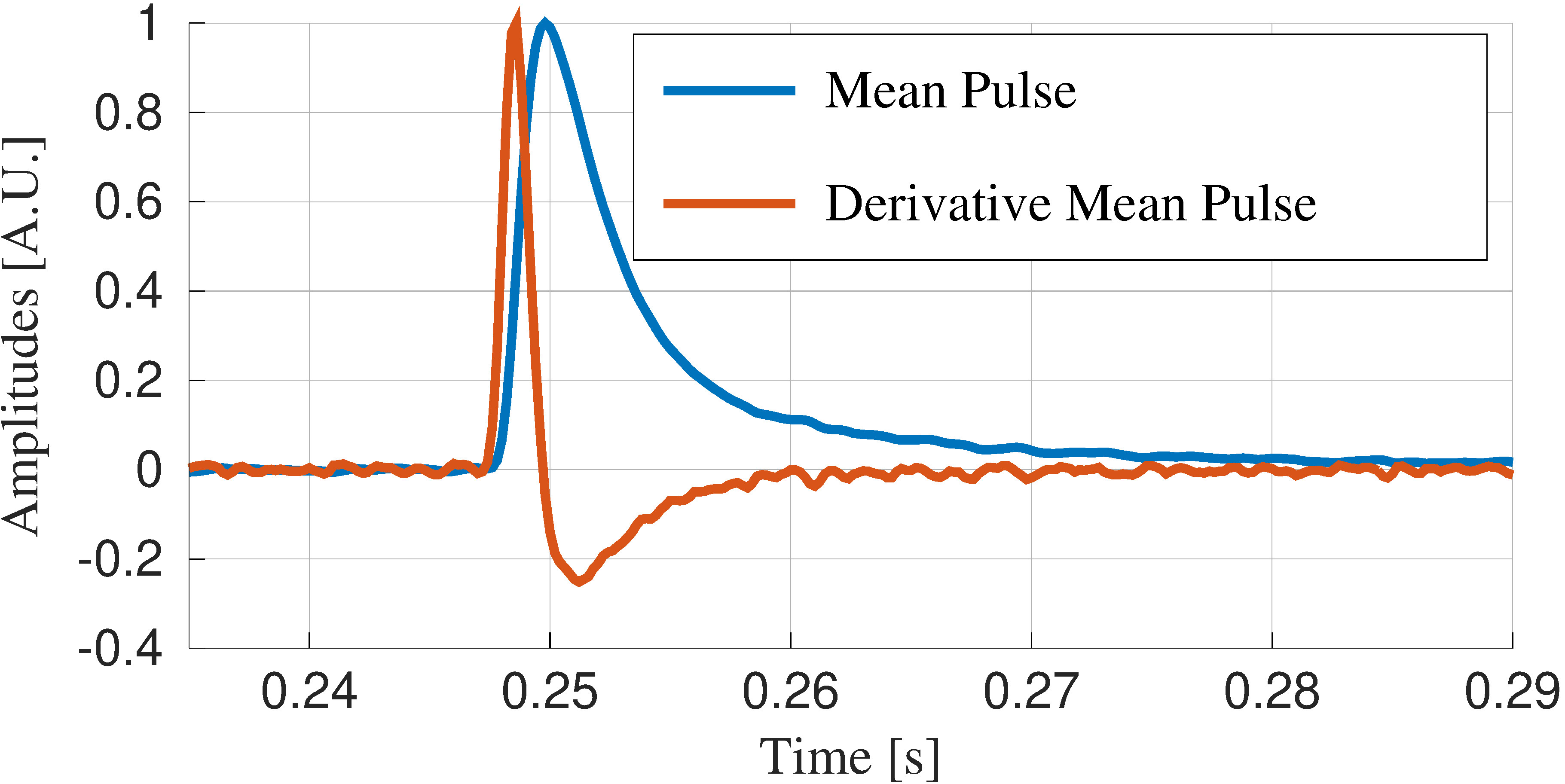}
\caption{\label{fig:lightpulse} Mean, normalized LED pulse $m(t)$ and its derivative $\frac{d}{dt}m(t)$. A 500~ms time-window was used for the Gatti-Manfredi optimal filter analysis.}
\end{figure}

To optimally reconstruct the energy of an event over the noise by exploiting the maximum of a signal pulse we employed the optimal filtering algorithm reported in \cite{Gatti:1986}. Few elements are required: 1) the typical average baseline power spectrum of the measurements, obtained by using data streams where no pulses are present (more than 200 baseline windows); 2) a mean pulse $m(t)$, obtained by averaging 100 pulses with the lowest used $\Delta$t of 50 $\mu$s. This value being much shorter than the rise-time we can consider these events as singles (Figure \ref{fig:lightpulse}).

We then built a transfer function $H(f)$ accordingly and filtered each pulse in the frequency domain; the maximum $A$ of the filtered pulse $s_f$(t) in the time domain is then, by construction, the pulse amplitude (energy) estimator which provides the best signal-to-noise ratio (S/N).

In order to evaluate the difference in shape of each pulse (pile-up) with respect to the average pulse (single), every filtered signal is compared in the time-domain with the filtered average pulse $m_f$(t) by fitting with $s_f$(t)${ }={ }$$A_{fit}{ }\cdot{ }$$m_f$(t+$\delta$t); the $\chi^2$ is minimized with respect to the $A_{fit}$ and $\delta$t. The minimization returns a second pulse amplitude estimator $A_{fit}$. 


For each signal we can determine a pulse shape parameter discriminator as PSD = $\frac{\textrm{$A$}}{\textrm{$A$}_{fit}}$: it will have value of 1 for pulses identical to that of the average pulse, lower if the shape is different as for pile-up events, and measures how much a signal differs from the reference pulse.

Particular attention was devoted to compare the pulse shape of particle-induced scintillation events with LED-generated ones; to this end we analyzed $^{232}$Th-source calibration run (48 hours) and selected the $^{208}$Tl (2615 keV $\gamma$s) scintillation events. We observed very minor differences in the shape, whose effect is negligible with respect to the scope of this work.

\begin{figure}[!h]
\centering
\includegraphics[width=0.45\textwidth]{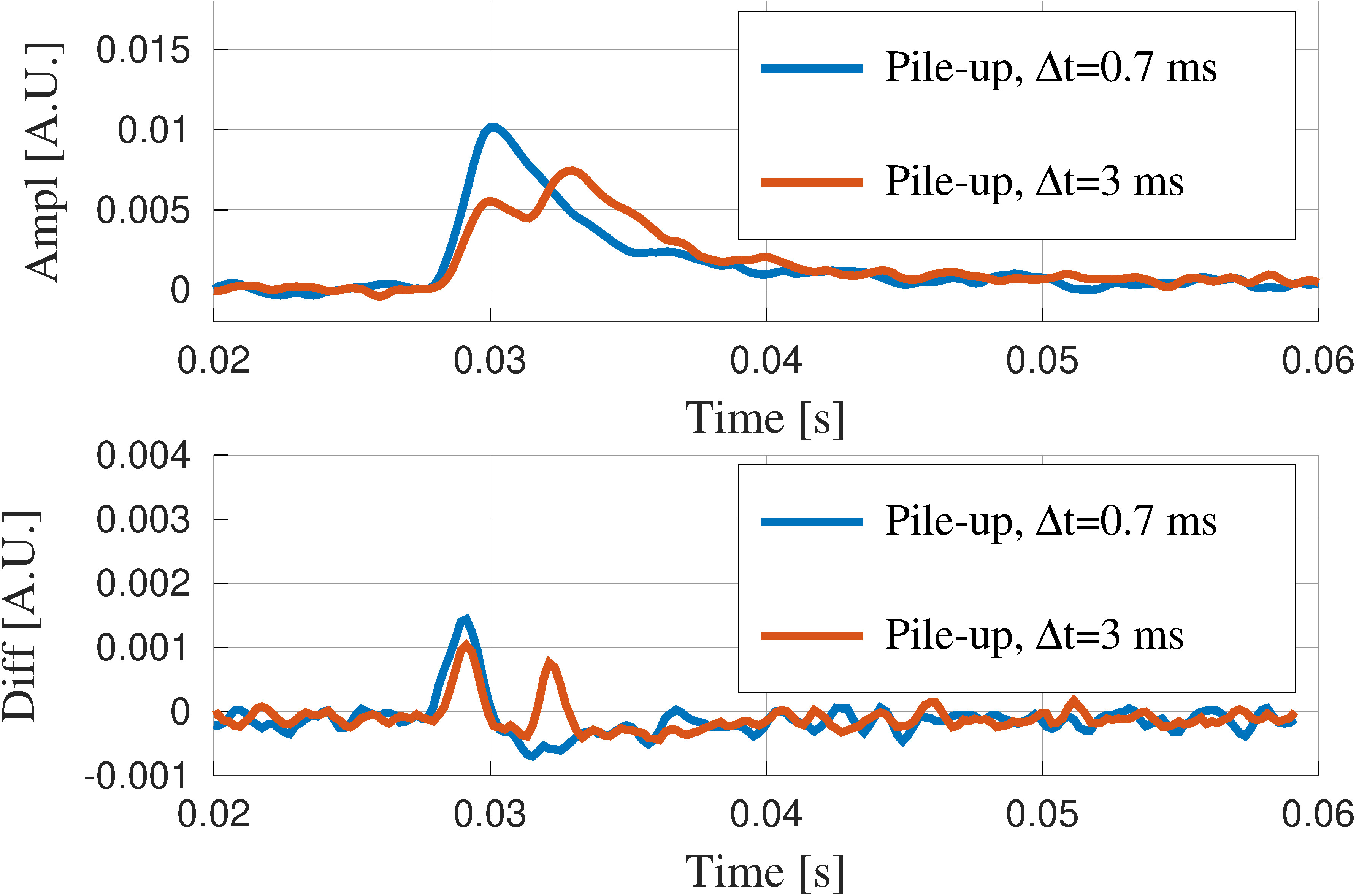}
\caption{\label{fig:exemple} Example of two 0.45 keV pile-up pulses (top) and derivatives (bottom, time-shifted to have the maximum centered at t=0.03 s) recorded by the NTL-LD at V$_e{ }={ }$50~V, for a pile-up time $\Delta$t of 3 ms (red) and $\Delta$t${ }={ }$0.7 ms (blue), respectively. Our pulse shape parameter algorithm utilizes the derivative of the signals and allows to reject 95\% of pile-up pulses with $\Delta t{ }={ }$0.7~ms.}
\end{figure}

To better distinguish signals differing in the rise-time part it is convenient to consider the derivative of the signal instead of the signal itself. In the following we will define as PSD$_{sig}$ the shape parameter based on the signal itself, and PSD$_{der}$ on the derivative of the signal. Figure~\ref{fig:lightpulse} shows the average signal and its derivative, whose maximum corresponds to the maximum slope of the signal in its rising part. With simple, geometrical consideration we can determine that a single event has a derivative maximum larger than that of a pile-up event at the same energy. Figure~\ref{fig:exemple} shows an example of two pile-up signals separated by 0.7 ms and 3 ms, together with their derivatives.

\begin{figure}[!h]
    \centering
    \includegraphics[width=0.45\textwidth]{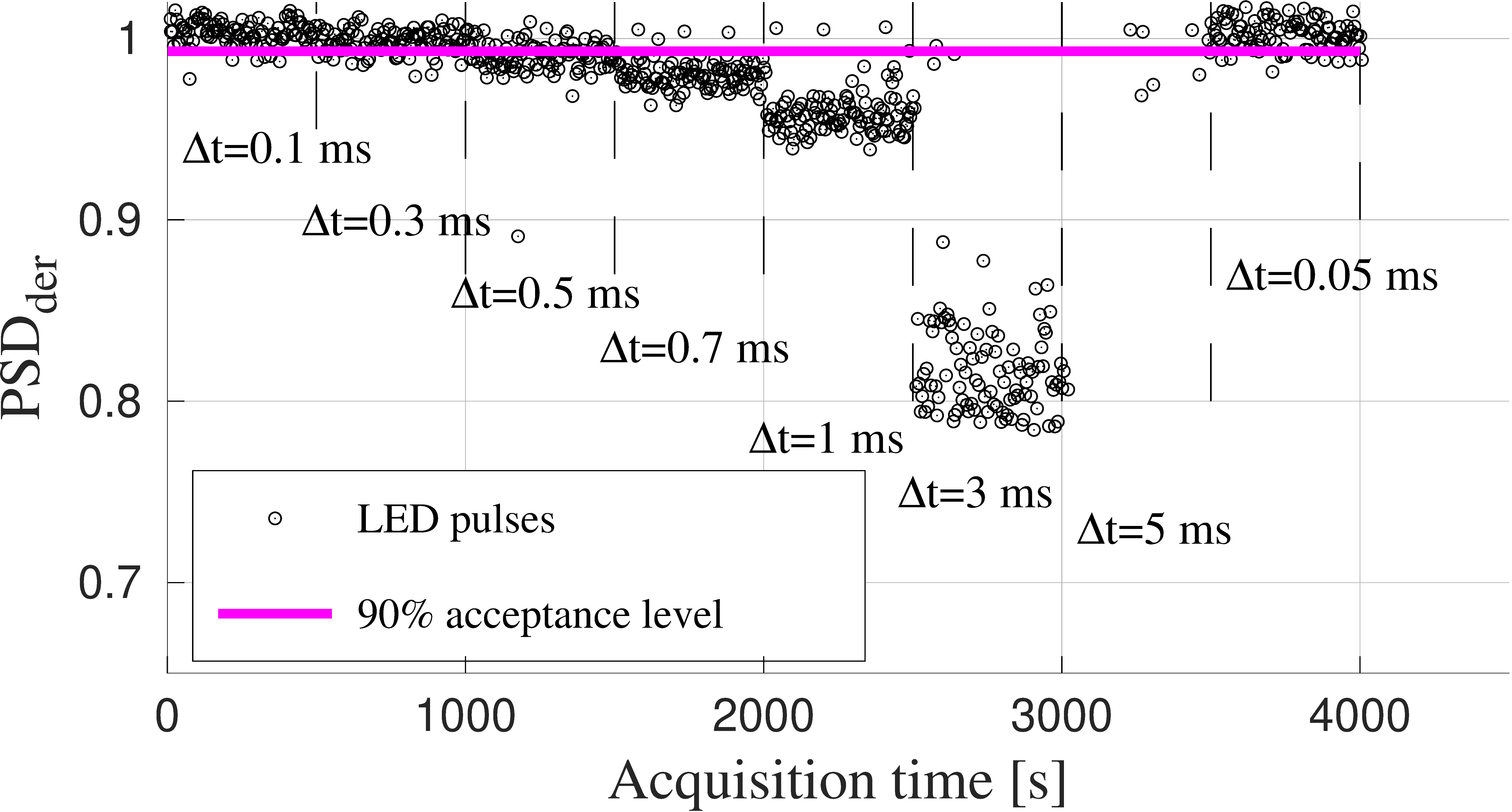}
    \caption{Scatter-plot of the PSD$_{der}$ as a function of the acquisition time (the detector is biased at V$_e{}={}$50~V). Every 500 s the $\Delta$t between the two pile-up pulses is changed, according to the pattern reported in the text. The distribution allows to trace the pile-up rejection curve with respect to $\Delta$t.}
    \label{fig:experiment}
\end{figure}
\begin{figure}[!h]
\centering
\includegraphics[width=0.4\textwidth]{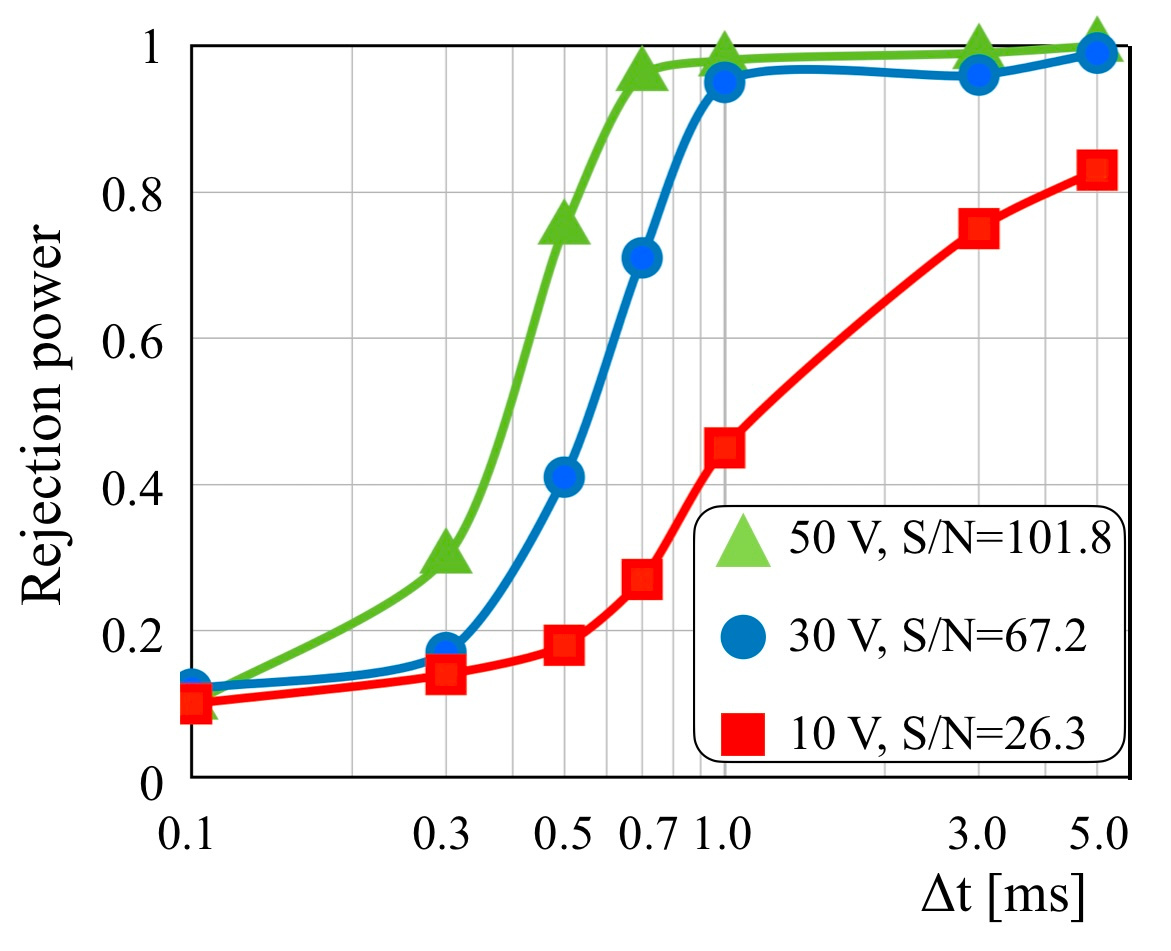}
\caption{\label{fig:rejectionpowercurve} Experimental rejection curves traced for several V$_e$ (continuous lines are used to guide the eye). The S/N on the filtered signal $s_f(t)$ spans from 9.6 at V$_e=$~0~V to 101.8 at V$_e=$~50~V,  corresponding to an effective NTL gain of about 10.}
\end{figure}

From now on we will work exploiting PSD$_{der}$ instead of PSD$_{sig}$.
Figure~\ref{fig:experiment} reports the PSD$_{der}$ as a function of the running time for a full measurement acquired at V$_e${ }={ }50~V and LED pulses injected according to the $\Delta$t time pattern as defined in Section~\ref{subsec:meas}. 
A PSD$_{cut}$ value, shown as a pink horizontal line in Fig.~\ref{fig:experiment}, is determined and adjusted such as to keep 90\% of single-pulse events ($\Delta$t{ }=~0.05~ms). For each group of pile-up pulses (100 consecutive) the number of events $N_{rej}$ with \break PSD$_{der}<$ PSD$_{cut}$ is counted and compared to the total number $N_{inj}$ of events injected. The rejection factor is defined as $r_{\Delta t}=N_{rej}/N_{inj}$ and the pile-up rejection power curves are traced for each set of measurements.
Figure~\ref{fig:rejectionpowercurve} shows all the experimental results together\footnote{For V$_e=0$~V the pulses are almost buried in the signal noise; the corresponding rejection power being far away from the goal.}; for completeness, for each set the S/N ratio is also given, which shows that the rejection capabilities increases with increasing S/N ratio, as also reported in \cite{Chernyak:2014,Chernyak:2017}, and expected from signal processing in general. 

\section{Simulations }\label{sec:sim}
\subsection{Synthetic data generation, analysis and results}
In order to reproduce, control and understand our pile-up experimental data, we adopted a method which permits to generate synthetic data, indistinguishable from the real, measured data \cite{Helis:2021}. The method is hereafter described.

Starting from a record of data (streaming) we proceed by: 
\begin{itemize}
    \item[1)] constructing an average signal $m(t)$ in the exact same condition as previously performed using the experimental data (see above); 
    \item[2)] summing the average signal re-scaled to an amplitude A1 with an another average signal of re-scaled amplitude A2, but time-shifted of $\Delta$t; 
    \item[3)] inserting the previously constructed pile-up pulses to the experimental data streaming (the full measurement was taken into account and we made sure that in the selected region no pulses were present).  
    \end{itemize}

\begin{figure}[!h]
\centering
\includegraphics[width=0.45\textwidth]{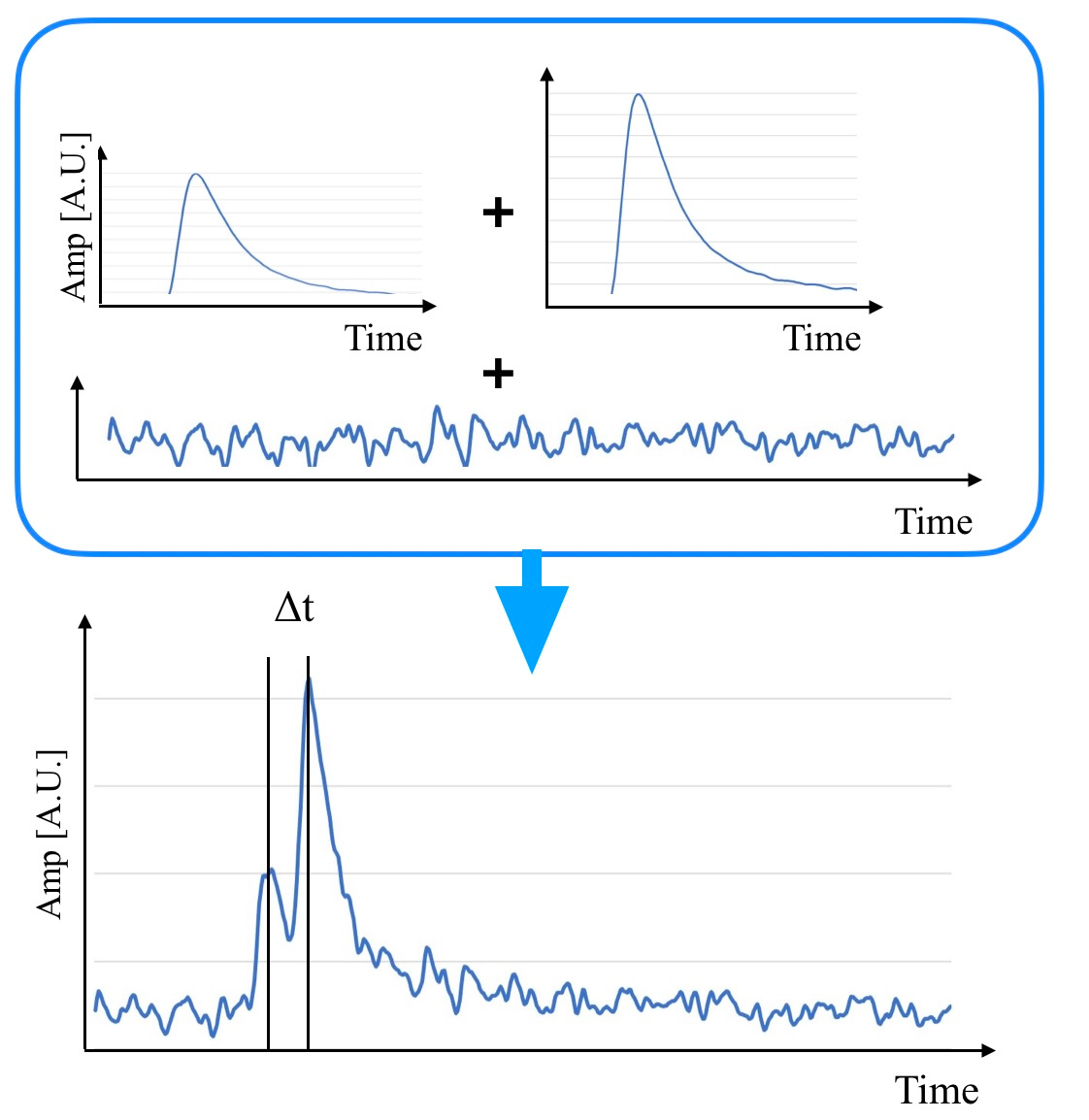}
\caption{\label{fig:synthetic} Two average signals of different amplitudes are summed together with a time difference of $\Delta$t and with a noise sample to give the synthetic data shown in the bottom of the figure.}
\end{figure}

    \noindent The full procedure is sketched in Fig.~\ref{fig:synthetic}.  \hfill
   \break

We built the synthetic data streaming conforming with our experimental measurements, namely equal amplitudes of the two signals, identical $\Delta$t pattern and the same number of pile-up events and repetition rate. The synthetic data are then elaborated using the same analysis pipeline of the experimental data. The pile-up rejection values $r_{\Delta t}$ are hence extracted as described in the previous section.

\begin{figure}[!h]
\centering
\includegraphics[width=0.45\textwidth]{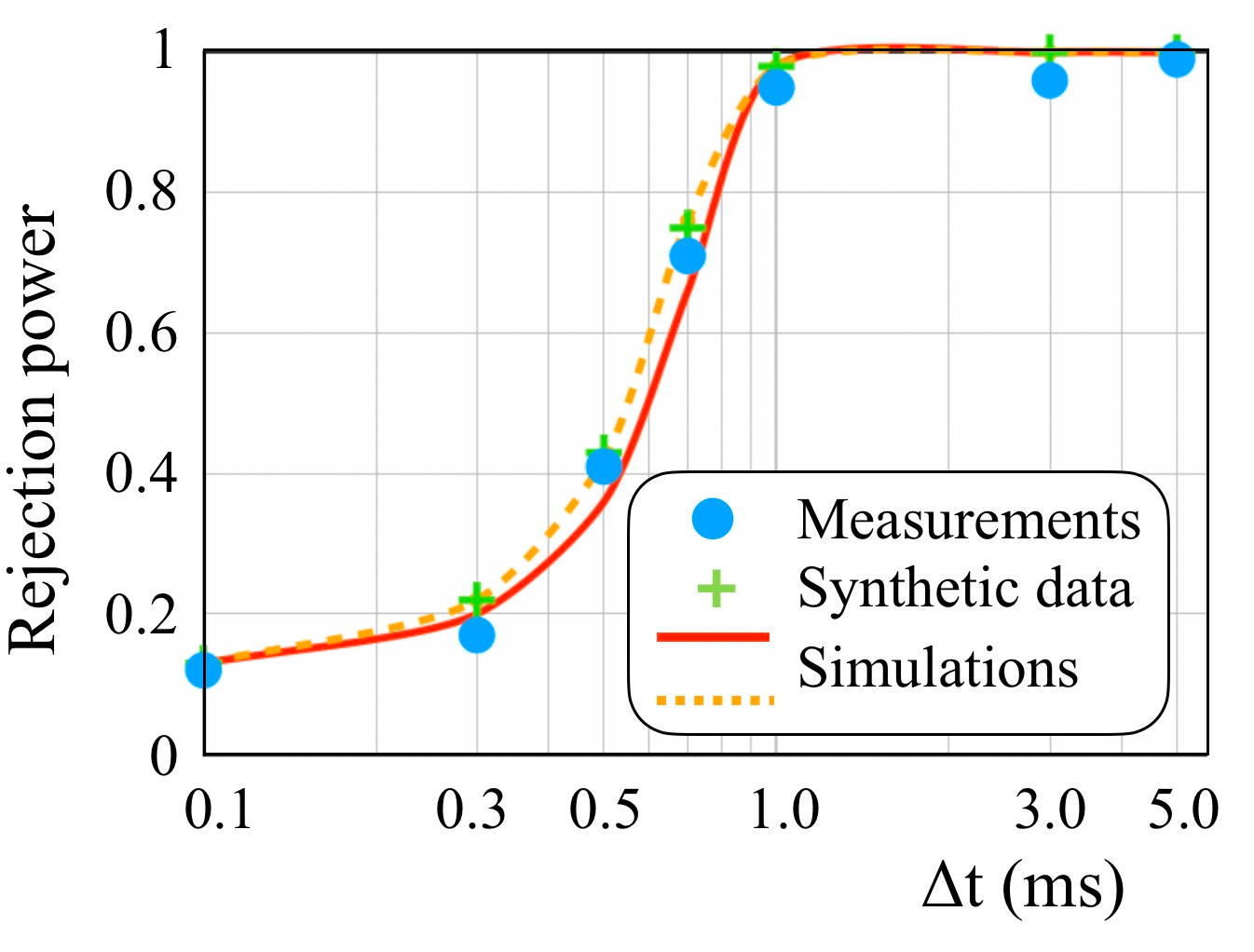}
\caption{\label{fig:rejvsdt} Rejection power curves versus $\Delta$t for an electrode bias V$_e${ }={ }30~V. Blue dots are from the measurements (see Fig.~\ref{fig:rejectionpowercurve}), green crosses are from the synthetic data, red plain curve is from the full simulation, $i.e.$ A1$\&$A2 $\in$ $2\nu2\beta$-spectrum, and orange dashed curve is from the approximate simulation with \break  A1${ }={ }$A2. Error bars are within the symbols. }
\end{figure}

To benchmark the quality of the synthetic data we compared in Fig.~\ref{fig:rejvsdt} the synthetic pile-up rejection performances with the experimental one.
An excellent agreement is observed in the whole range, within the uncertainties\footnote{Bayesian statistics calculation was used to infer the error bars on the experimental data and are within the symbols. The variance for a having $k$ out of $n$ occurrences reads: \break $V=\frac{(k+1)(k+2)}{(n+2)(n+3)}-\frac{(k+1)^2}{(n+2)^2}$. We finally have 68\% within $k\pm\sigma$ where $\sigma=\sqrt{V}$. For a rejection of 0.1 ($\frac{k}{n}=0.1$) and a number of occurrences of 100, $\sigma$${}={}$0.03. They are equal as for the synthetic simulation as the number of occurrences is the same, namely 100 events for each $\Delta$t \cite{Ullrich:2007}.}, between the measurement and the synthetic data.

In the next section we will show how we performed wider synthetic data simulations by modifying the pile-up time step, number of occurrences, and A1/A2 ratio to best comply with a real double-beta decay experiment.

\subsection{Refined synthetic pile-up simulation}\label{fullsim}
First of all we refined our synthetic data by producing pile-up events with a $\Delta$t running from 0 to 3~ms with a time step of 0.1~ms, and injecting 1800 events for each, assuming equal amplitudes for the two signals (A1${}={}$A2). This produces the orange curve shown in Fig.~\ref{fig:rejvsdt}.

\begin{figure}[!h]
\centering
\includegraphics[width=0.4\textwidth]{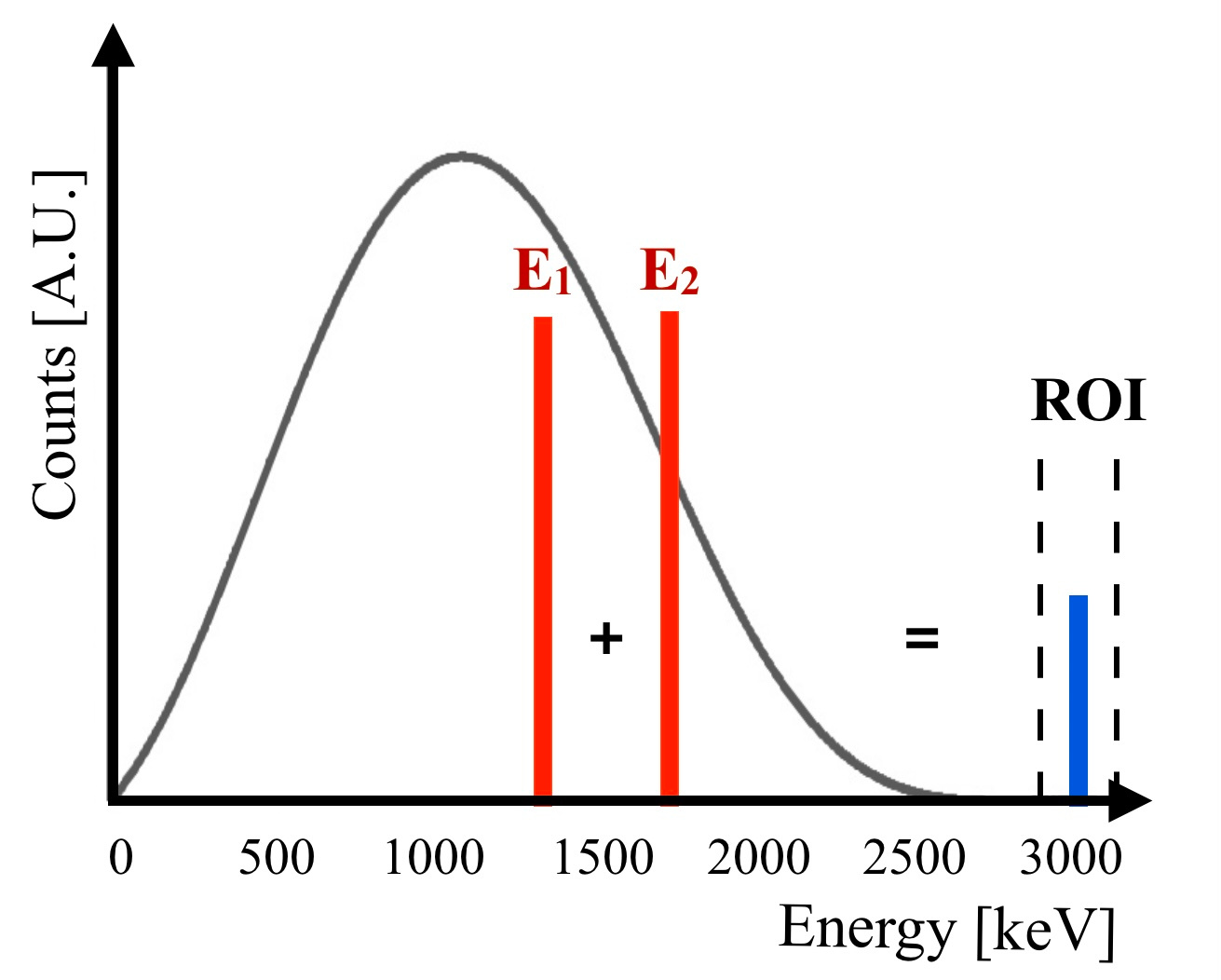}
\caption{\label{fig:dbds} Energy spectrum of $^{100}$Mo 2$\nu$2$\beta$ decay. The two red bars correspond  to the energies of randomly picked up values which sum up to an energy in the ROI of $^{100}$Mo 0$\nu2\beta$ decay.}
\end{figure}

However, a realistic synthetic simulation (called hereafter full simulation) must take into account that the amplitudes of the two signals, A1 and A2, should comply with the $2\nu2\beta$ energy distribution \cite{Chernyak:2012}. Therefore, pile-up events are constructed by picking up twice randomly inside this energy distribution and summing them up with a $\Delta$t ranging from 0 to 3~ms. We finally select those events having an energy in the ROI, {\it i.e.} 3034 keV $\pm$ 50 keV (Fig.~\ref{fig:dbds}).
The full simulation pile-up rejection curve is shown in Fig.~\ref{fig:rejvsdt} as red solid line. 
It should be noticed that we expect a lower pile-up rejection efficiency than for the case where pulses have identical amplitudes (orange curve). This is due to the fact that for A1${}<<{}$A2 (A1${}>>{}$A2) pile-up events may be barely distinguishable from single-pulses and may not be rejected. This is precisely what is observed in Fig.~\ref{fig:rejvsdt}  as the red plain curve (A1${}\neq{}$A2) is below the orange dashed one (A1${}={}$A2). The full simulation and the approximate simulation returning almost the same rejection power curves motivates the simplified approach employed in the experimental work.

\subsection{Background index extraction}
To extract the background index resulting from pile-up events from these simulations we need to compute the probability, $\epsilon_{Q_{\beta\beta}}$($\Delta$t), to obtain an event at 3034 keV resulting from a pile-up of two events coming from the 2$\nu$2$\beta$ decay for different values of $\Delta$t between the two single events. 

\begin{figure}[!h]
\centering
\includegraphics[width=0.45\textwidth]{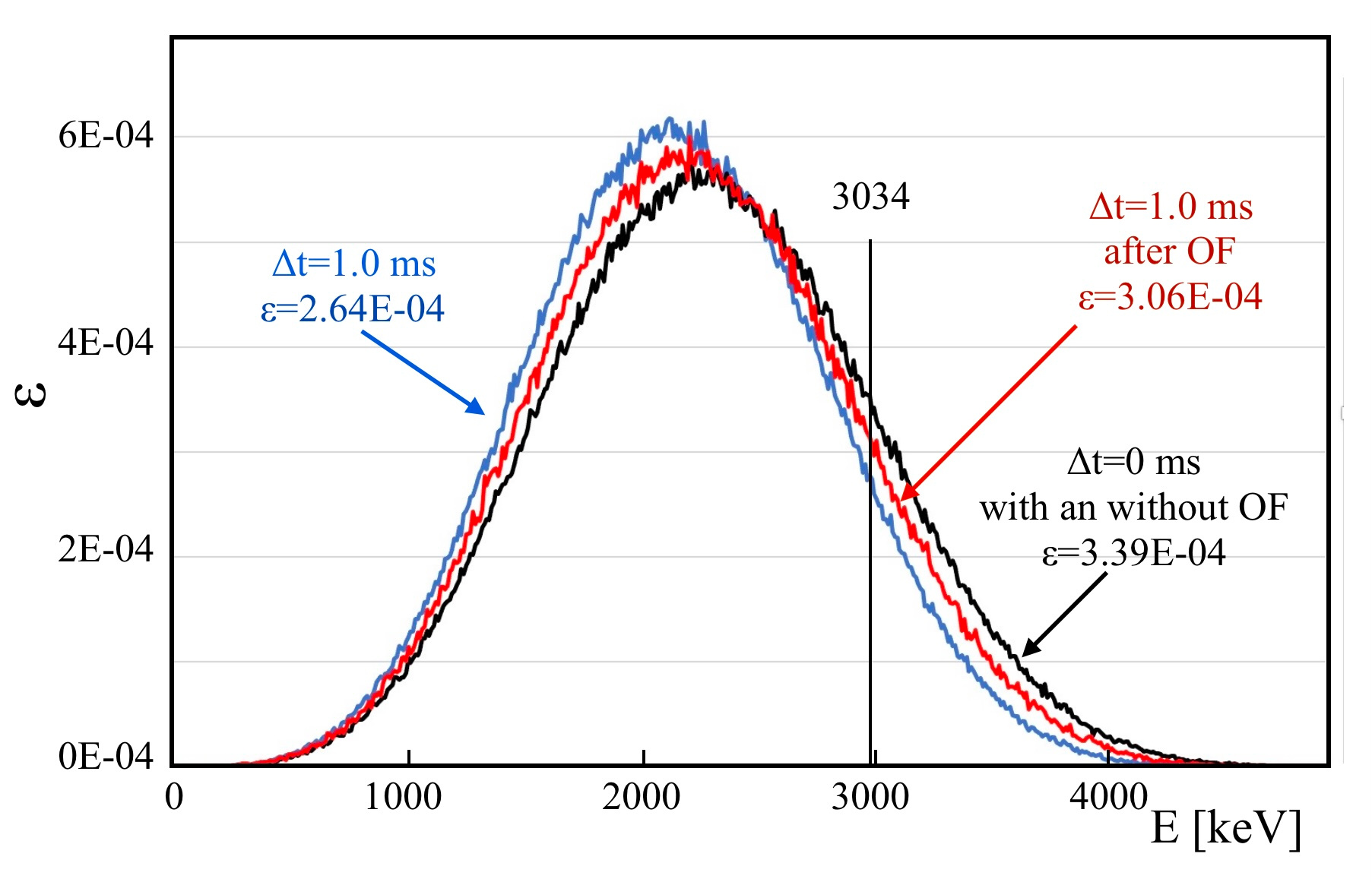}
\caption{\label{fig:BIDT} Distributions of the sum of two 2$\nu$2$\beta$ decays separated by $\Delta$t equal 0 and 1 ms after the optimal filtering procedure. The probability $\epsilon$ is given at 3034 keV.}
\end{figure}

These probabilities are shown in Fig.~\ref{fig:BIDT} where the distributions of the amplitude of the pile-up events are presented for $\Delta$t${ }={ }$0 and 1~ms. The rightmost distribution corresponds to $\Delta$t${ }={ }$0 ms and does not depend on the signal shape since it is identical to the initial average signal. Hence the amplitude of the signal does not vary after the optimal filtering whereas for $\Delta$t{ }$\neq$ {}0 it does. The presented distributions were obtained for an analytical signal of 1.2~ms rise-time, the description of which is presented hereafter. The leftmost curve (blue) corresponds to the amplitude of the pile-up events for $\Delta$t $=1$~ms and the middle (red) one corresponds to its distribution, slightly distorted by the optimal filtering procedure.

We then calculated the background indexes associated to pile-ups of $^{100}$Mo 2$\nu$2$\beta$ events for every simulated configuration (Table \ref{tab:backind})
as follows:
\begin{equation}
BI=\epsilon_{Q_{\beta\beta}}\times (1-r_{\Delta t})\times \tau\times(\frac{ln2\cdot N_{^{100}Mo}}{T_{1/2}^{2\nu2\beta}})^2 \quad [\mathrm{ckky}]
\end{equation}

\noindent where $\tau$ is the time step (here 0.1 ms as in the simulations), $N_{^{100}Mo}$ the number of $^{100}$Mo nuclei and $T_{1/2}^{2\nu2\beta}$ is the half-life of the $^{100}$Mo 2$\nu$2$\beta$ decay. 

In the next section we study how the signal rise-time changes the pile-up rejection efficiency.

\subsection{Signal rise-time and pile-up rejection efficiency}

It is of major interest to investigate and predict how the light detector rise-time $\tau_r$ will affect the pile-up rejection capability, combined with the S/N ratio. To this end, we generated an analytical average pulse, which is profiled by adding a Gaussian function, to account for the rising part of the signal, with a sum of exponential functions, to account for the decay part of the signal. 

\begin{figure}[!h]
\centering
\includegraphics[width=0.45\textwidth]{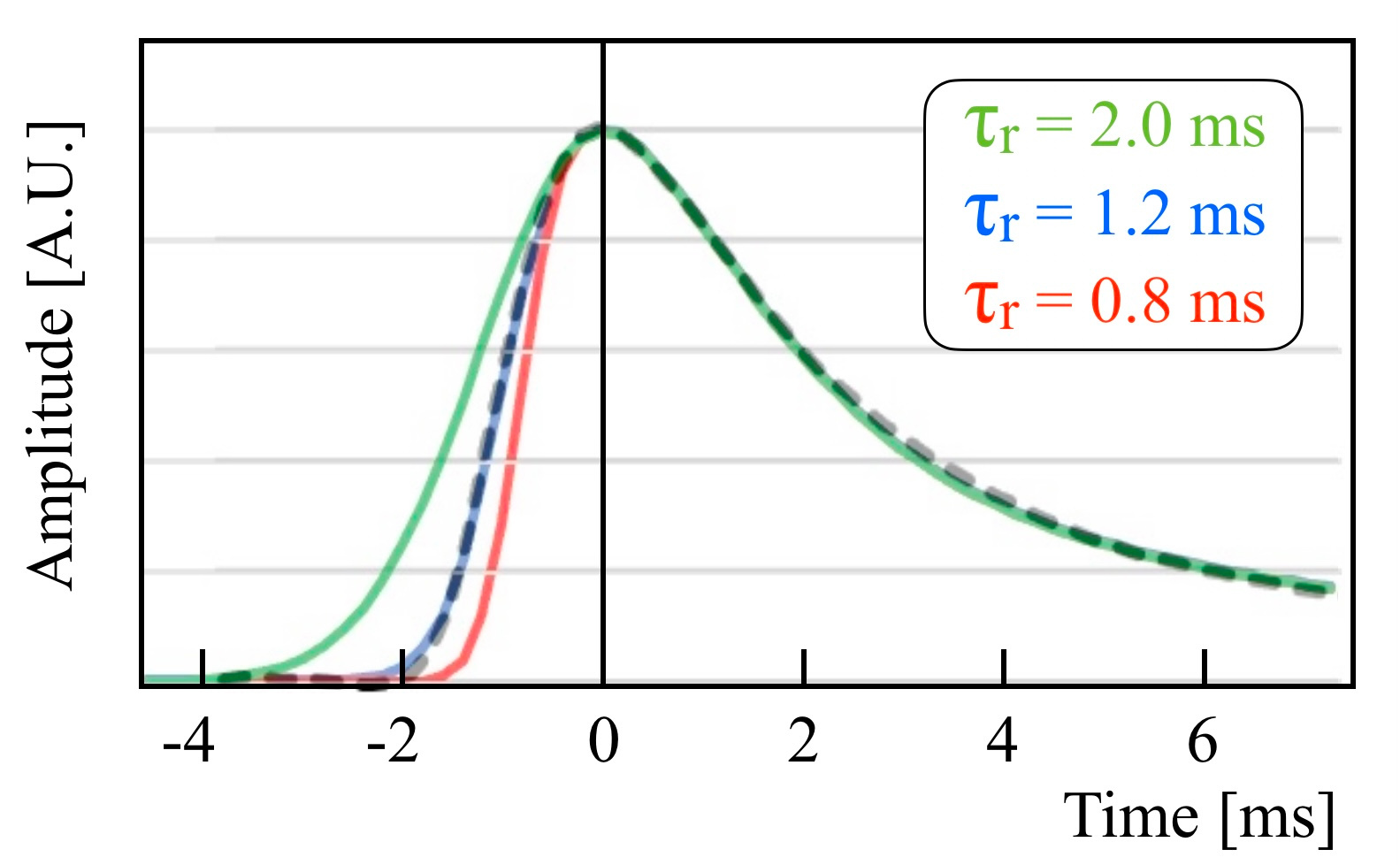}
\caption{\label{fig:anasig} Average signals coming from the fit as described in the text for rise times equal to 0.8, 1.2 and 2.0 ms. The gray dashed line corresponds to the average signal for 1.2 ms rise time. }
\end{figure}

Figure \ref{fig:anasig} shows three analytic average pulses with different rise-time values; even though the blue curve ($\tau_r$ = 1.2~ms) does reproduce fairly well the experimental average pulse $m(t)$ (dashed gray line) extracted from the measurements, all the simulations with these rise-time were performed with the experimental average pulse.
Analytical pulses enable to make faster/slower average pulses in a simple manner, to produce different synthetic simulation data and extract, accordingly, the power rejection curves $r_{\Delta t}=f(S/N,\tau_r)$. The variation of the pile-up rejection power curves with respect to the pulse decay time is less important and therefore is not taken into account in this study.

\begin{figure}[!h]
\centering
\includegraphics[width=0.45\textwidth]{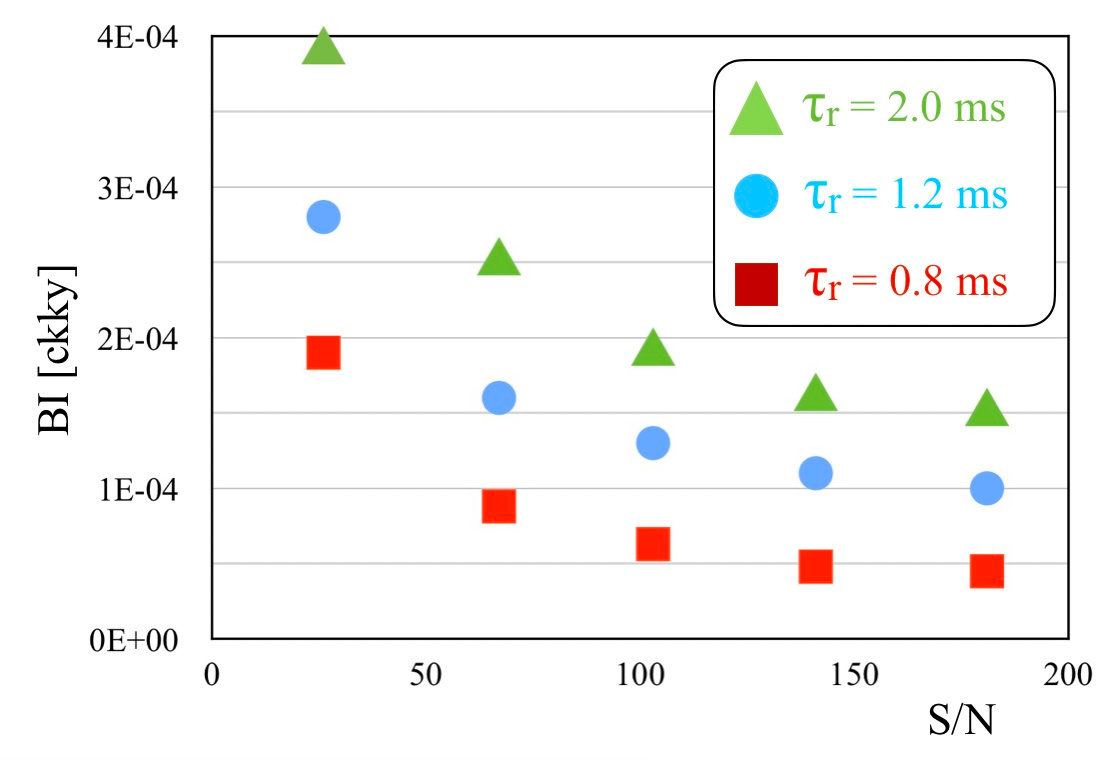}
\caption{\label{fig:magnificient} Background index associated to $^{100}$Mo 2$\nu$2$\beta$ decay events, after a PSD$_{der}$ selection, for three values of the signal rise-time and several S/N ratios (Table~\ref{tab:backind}). The simulations of the blue circles corresponding to $\tau_r{ }={ }$1.2 ms were done with the average experimental signal whereas the two others, \break $\tau_r{ }={ }$2 ms   and $\tau_r{ }={ }$0.8 ms, used the analytical signals described in Fig.~\ref{fig:anasig}.  }
\end{figure}

We have performed full simulations with the pulses of Fig.~\ref{fig:anasig}, varying for each the S/N ratio accordingly; results are displayed in Fig.~\ref{fig:magnificient}. It should be noticed that the S/N achieved correspond to the 1.2 ms rise-time signal (as in the experiment) and for NTL-LD electrode biases V$_e { }={ }$10,\ 30,\ 50~V, the three left round blue dots of Fig.~\ref{fig:magnificient}. The S/N corresponding to 70 and 90~V were extrapolated.

The method here presented is a powerful tool to predict the pile-up rejection power and hence the pile-up background index achievable in the ROI, by simply combining few, generic detector specification as S/N ratio  and rise-time. 

\subsection{Approximate evaluation of the background index}
As we performed the full simulations for every bias, we were able to determine the $\Delta$t value corresponding to 50\% pile-up rejection ($\Delta$t$_{50\%}$). This latter was calculated  using a linear interpolation between the two consecutive $\Delta$t leading to rejection below and above 50\%. Corresponding values are reported in Table \ref{tab:backind} as well as the background index.
The background index associated to pile-ups of $^{100}$Mo 2$\nu$2$\beta$ events in a 280~g bolometer, 95\% $^{100}$Mo-enriched and using the PSD$_{der}$ parameter can be approximated as:
\begin{equation}
BI \approx \Delta t_{50\%}\times2.75\cdot10^{-4} \quad [\mathrm{ckky}],
\end{equation}
where $\Delta$t$_{50\%}$ is expressed in ms. 

Because of the high slope of the rejection power curve around $\Delta t_{50\%}$, the above equation gives a better feeling of the BI that can be achieved than by simply extrapolating from the $\Delta t$ value where 90\% rejection is obtained as often expressed \cite{Armatol:2021b}. With this rule of thumb hereby expressed, in order to obtain $5\cdot10^{-5}$~ckky \cite{CUPID:2019} we should reach 50\% rejection for signal separated by  $\Delta$t = 0.18 ms. 

\begin{table*}[]
    \centering
    \caption{Background index values for different S/N ratio values and signal rise-time as shown in Fig.~\ref{fig:magnificient}.  $\Delta$t$_{50\%}$ are the $\Delta$t  corresponding to 50\% of pile-up rejection for signals of $\tau_r{ }={ }$1.2 ms. }
    \label{tab:backind}
    \begin{tabular}{c|c|c|c|c|c}
         Bias & S/N & \multicolumn{3}{|c|}{BI [in 10$^{-4}$ ckky]} & $\Delta$t$_{50\%}$ [ms] \\
         $[$V$]$ & ~ & for  $\tau_r{ }={ }$2 ms  &  for $\tau_r{ }={ }$1.2 ms  & for $\tau_r{ }={ }$0.8 ms &  for $\tau_r{ }={ }$1.2 ms \\
         \hline
         10 & 26  & 3.9 & 2.7 & 1.9 & 0.91 \\
         30 & 67  & 2.5 & 1.6 & 0.88& 0.58 \\
         50 & 103 & 1.9 & 1.3  & 0.63& 0.47 \\
         70 & 141 & 1.6 & 1.1 & 0.48 & 0.40 \\
         90 & 181 & 1.5 & 1.0 & 0.45 & 0.37 \\
         
    \end{tabular}

\end{table*}

\section{Conclusions}
Some future bolometric experiments, like  CROSS and CUPID, rely on a detector technology based on NTD-equipped scintillating bolometers to search for the $0\nu2\beta$ of $^{100}$Mo. Due to the poor time resolution of the main heat bolometer, randomly-coincident (pile-up) $^{100}$Mo  $2\nu2\beta$ decay events will eventually represent the main source of background in the next-generation, large-scale, 0$\nu$2$\beta$ decay searches based on this technology. 
To circumvent this problem we have operated a Li$_2$MoO$_4$ scintillating bolometer in combination with a Neganov-Trofimov-Luke light detector. Thanks to the fast time response and the enhanced S/N ratio performances of this latter, we investigated the rejection of pile-ups from single-pulse events utilizing the mere scintillation signal. We studied the pile-up rejection performance of the setup in a validation run  hosted in the CROSS cryogenic underground facility. We demonstrated for first time with an experimental pile-up measurement coupled with a simulation,  that it is possible to reject randomly-coincident events in massive Li$_2$$^{100}$MoO$_4$ cryogenic scintillating bolometers, as those of CROSS and CUPID experiments, down to background indexes of $\sim10^{-4}$ ckky (at the Q$_{\beta\beta}$ of $^{100}$Mo), via the scintillation signal.

Together with the experimental work, we presented (i) an approximate and full simulation  method to reconstruct, synthetically, randomly coincident events from the $2\nu2\beta$ decay mode, (ii) a new pulse-shape discrimination algorithm based on a derivative, optimal-filtering signal processing technique capable to provide superior pile-up rejection efficiencies.
The experimental and simulation pile-up rejection results have been compared: we observed an excellent agreement between them. As we have shown that there is a strong dependence with the signal rise-time, it is strongly recommended that we use the real signal such as the one coming from calibrations to extract the average signal to be used in the simulation to predict the achievable background index.

With respect to CUPID, the CROSS detectors foresee to have the additional capability to reject surface events \cite{Bandac:2021}. The reduction of the random-coincidence background enables to fully exploit the CROSS surface sensitive technology. In fact, the background induced by beta particles emitted by the radioactive contamination of the passive materials facing the detectors is a sub-dominant contribution of the background in the ROI that will emerge only after the mitigation of the random-coincidence component.

New investigations are ongoing to further enhance the light detector S/N ratio via higher NTL electrode bias, shorten the light detector rise-time and hopefully reach background indexes in the ROI of $5\cdot10^{-5}$~ckky, which would fully comply with the CUPID experiment goal.

\section*{Acknowledgments}
This work is supported by the European Commission (Project CROSS, Grant No. ERC-2016-ADG, ID 742345) and by the Agence Nationale de la Recherche (Project CLYMENE; ANR-16-CE08-0018; Project CUPID-1; ANR-21-CE31-0014, ANR France). We acknowledge also the support of the P2IO LabEx (ANR-10-LABX0038) in the framework ''Investissements d'Avenir'' (ANR-11-IDEX-0003-01 – Project ''BSM-nu'') managed by ANR, France.

\end{document}